\documentclass[]{bytedance_seed}

\usepackage[toc,page,header]{appendix}

\usepackage{minitoc}
\usepackage{amsmath,amssymb}
\usepackage{booktabs} 
\usepackage{xcolor}   
\usepackage{tabularx} 
\usepackage{graphicx} 
\usepackage{colortbl} 
\usepackage{siunitx} 
\usepackage{makecell}      
\usepackage{fancyvrb}
\usepackage{epigraph}

\newcommand{\code}[1]{\colorbox{black!10}{\texttt{#1}}}

\title{Understanding by Reconstruction: Reversing the Software Development Process for LLM Pretraining}

\author[1,2 *]{Zhiyuan Zeng}
\author[3, *]{Yichi Zhang}
\author[1, *]{Yong Shan}
\author[1, *]{Kai Hua}
\author[3]{Siyuan Fang}
\author[4]{Qian Liu}
\author[3]{Jiaheng Liu}
\author[6,3]{Haozhe Wang}
\author[2]{Yining Zheng}
\author[1]{Ming Ding}
\author[1]{Ke Shen}
\author[1 \dagger]{Ge Zhang}
\author[1, \dagger]{Wenhao Huang}
\author[2,5 \dagger]{Xipeng Qiu}

\affiliation[1]{ByteDance Seed}
\affiliation[2]{Fudan University}
\affiliation[3]{M-A-P}
\affiliation[4]{ByteDance TikTok}
\affiliation[5]{Shanghai Innovation Institute}
\affiliation[6]{The Hong Kong University of Science and Technology}

\contribution[*]{Core Contributors}
\contribution[\dagger]{Corresponding authors}

\abstract{
While Large Language Models (LLMs) have achieved remarkable success in code generation, they often struggle with the deep, long-horizon reasoning required for complex software engineering. We attribute this limitation to the nature of standard pre-training data: static software repositories represent only the terminal state of an intricate intellectual process, abstracting away the intermediate planning, debugging, and iterative refinement. To bridge this gap, we propose a novel paradigm: understanding via reconstruction. We hypothesize that reverse-engineering the latent agentic trajectories—the planning, reasoning, and debugging steps—behind static repositories provides a far richer supervision signal than raw code alone. To operationalize this, we introduce a framework that synthesizes these trajectories using a multi-agent simulation. This process is grounded in the structural realities of the source repositories (e.g., dependency graphs and file hierarchies) to ensure fidelity. Furthermore, to guarantee the logical rigor of the synthetic data, we employ a search-based optimization technique that iteratively refines the Chain-of-Thought (CoT) reasoning to maximize the likelihood of the ground-truth code. Empirical results demonstrate that continuous pre-training on these reconstructed trajectories significantly enhances Llama-3-8B’s performance across diverse benchmarks, including long-context understanding, coding proficiency, and agentic capabilities.}

\date{\today}
\correspondence{Zhiyuan Zeng at \email{cengzy23@m.fudan.edu.cn}, Ge Zhang at \email{zhangge.eli@bytedance.com}}

\begin{document}
\maketitle


\section{Introduction}
The remarkable success of Large Language Models (LLMs) can be viewed as a modern validation of Richard Feynman's famous dictum: ``\textit{What I cannot create, I do not understand.}'' The dominant paradigm of generative pre-training \citep{floridi2020gpt,instruct-gpt} is built on this very principle---that the ability to generate text token-by-token serves as the proxy for understanding language. By learning to predict the next token, models internalize the syntax, semantics, and world knowledge embedded within vast corpora.

However, this ``understanding via generation'' paradigm faces a fundamental limit when applied to complex, long-horizon artifacts, such as substantial software repositories. A software repository, in its final form, is the terminal state of an intricate intellectual process. It is a highly compressed artifact where the ``computational steps'' of human reasoning---the requirement analysis, architectural planning, trial-and-error debugging, and iterative refinement---have been abstracted away. When we train models solely on this static code, we are essentially asking them to memorize the destination without showing them the map. Consequently, models often learn to mimic the surface-level structural patterns of the result rather than mastering the generative reasoning required to derive it. This explains why models that excel at generating short snippets often fail to grasp the deep, causal logic required to construct and maintain complex software systems \citep{pham2025swe}.

To bridge this gap, we propose that to truly understand a repository, a model should learn to reconstruct the process that created it. Our motivation is to reverse-engineer the latent agentic trajectory hidden behind static code~\cite{reer}. We hypothesize that by restoring the missing details of the generation process---explicitly expanding a static repository into a dynamic sequence of planning, reasoning, and execution steps---we can provide a far richer supervision signal than the raw code alone. This allows the model to learn not just what the code is, but why and how it was written, thereby aligning the training data more effectively with the model's next-token prediction objective.

To implement this data-centric philosophy, we developed a framework to synthesize these trajectories from existing high-quality open-source repositories. We treat the repository as the ground truth answer and simulate the problem-solving steps required to arrive there. Specifically, we employ a multi-agent simulation, where a main agent generates high-level requirements and implementation plans, while sub-agents are delegated to handle individual files. These agents utilize a ``Read'' tool to gather context and a ``Write'' tool to generate code. Crucially, to prevent the simulation from drifting, we inject structural ground-truth information---such as file hierarchies and dependency graphs extracted from the repository---to guide the agents, ensuring the synthesized trajectory faithfully reconstructs the target artifact.

While this reconstruction provides the ``missing steps'', the quality of the reasoning itself remains a variable. The initial CoT generated during simulation may be suboptimal. To address this, we introduce a search-based optimization technique to refine the thinking process. We posit that a high-quality thought ($z$) should maximize the likelihood of the correct code ($x$), formalized as maximizing $\log p(x|z)$. Drawing inspiration from tree-search algorithms \citep{qiu2024treebon, reer}, we decompose the trajectory into steps and iteratively sample refinements. We replace the original reasoning with refined thoughts only when they lower the perplexity of the target ground-truth code. This process polishes the synthetic trajectory, yielding a dataset that is not only causally complete but also logically rigorous.

We empirically validate our paradigm by continuously pre-training Llama-3-8B on our synthesized dataset. The results demonstrate that learning from these reconstructed trajectories leads to significant performance gains across diverse benchmarks, including long-context understanding, coding, reasoning, and agentic capabilities.

Our contributions are summarized as follows:
\begin{enumerate}
    \item We propose a novel paradigm for scaling LLM capabilities based on the principle of \textbf{understanding via reconstruction}. We argue that static repositories miss crucial generative details, and we introduce a method to reverse-engineer these latent agentic trajectories to provide richer supervision.
    \item We develop a multi-agent simulation framework that synthesizes these trajectories by grounding the generation process in the structural realities of source repositories, effectively converting static data into dynamic thinking and acting,.
    \item Experimental results show that Llama-3-8B, when pre-trained on our reconstructed data, achieves superior performance across benchmarks for long-context understanding, coding, reasoning, and agentic tasks.
\end{enumerate}
\section{Related Work}
\subsection{Reverse Reasoning in Pretraining Data}

The paradigm of ``reasoning recovery'' posits that logical capabilities are latent within pre-training data and can be activated through explicit structural modeling. Early efforts such as Quiet-STaR \citep{quietstar} internalize this process at the token level by training models to generate implicit rationales that minimize future-token uncertainty, effectively embedding an ``internal monologue'' within the model's latent space. Shifting toward structural optimization, BOLT \citep{bolt} learns latent reasoning for pre-training documents via an EM framework, systematically bridging the gap between raw text and logical derivation.  From a data-engineering perspective, Thinking Augmented Pre-training (TPT) \citep{tpt} prepends synthetic thinking trajectories to pre-training corpora, effectively reallocating computational budget toward logic-dense segments to enhance data efficiency. Most recently, REER \citep{reer} introduced a reverse-engineering approach for open-ended generation, utilizing perplexity-driven path searching to reconstruct the logical scaffolding behind high-quality reference answers.

Unlike existing approaches that recover isolated reasoning steps, our framework reconstructs a holistic agentic trajectory—integrating high-level architectural planning, file-level action sequencing, and iterative tool-use—thereby capturing the multi-dimensional generative process of entire repositories.

\subsection{Synthetic Agent Trajectories}
There are two primary methods for constructing agent trajectories in existing research.

The first method involves generating trajectories through agent exploration in real-world environments \citep{agent-trek,scale-explorer,team2025kimi,qwen-scale-env, wang2025code}. While this approach ensures the authenticity of the trajectories, it has significant drawbacks, including potentially expensive tool invocation costs and substantial engineering efforts required for environment setup and maintenance.

The second method places the agent in an environment simulated by LLM \citep{tool-alpaca,sim-rl,scale-agent-traj} or prompt LLM to generate an entire synthetic trajectory \citep{magv, reer}. The main advantage of this approach is its low cost. However, the resulting trajectories may suffer from extensive hallucinations generated by the LLM, compromising data reliability.

The trajectory synthesis method proposed in this paper draws inspiration from the second by using an LLM to generate both the tool calls and their corresponding outcomes. Although the synthesized trajectories may contain some noise, we ensure that their terminal state is a real repository, which serves as the ground truth.

\subsection{Synthetic Data for Coding}
The use of synthetic data has become a cornerstone in advancing the capabilities of LLMs for code-related tasks. A significant body of work focuses on generating instruction-following datasets. For instance, Magicoder \citep{wei2023magicoder} synthesizes user instructions for open-source code snippets to create a dataset aimed at enhancing the coding abilities of LLMs. Similarly, Code Alpaca \citep{codealpaca} employs the self-instruct methodology \citep{self-instruct} to generate a dataset of 20,000 code instructions. To improve the quality of these instructions, WizardCoder \citep{luo2023wizardcoder} introduces an evolutionary pipeline that progressively increases the complexity and diversity of the initial instructions.

Other research explores different forms of synthetic data. Case2Code \citep{shao2025case2code}, for example, collects a vast number of input-output test cases by executing existing programs and then generates new programs that satisfy these test cases. More recently, SWE-Synth \citep{pham2025swe} focuses on generating synthetic data for program bug fixing, which has proven effective in improving LLM performance on benchmarks like SWE-Bench \citep{jimenez2023swe}. The widespread adoption of such synthetic data in both the pre-training and post-training phases of modern Code LLMs, such as Qwen2 \citep{hui2024qwen2} and DeepSeek-Coder \citep{deepseek-coder}, underscores its critical importance~\cite{hui2024qwen2,wang2025emergent}.

While our work also contributes to the field of synthetic data for code generation, it diverges from previous efforts in two fundamental aspects. First, instead of augmenting isolated code snippets, we focus on augmenting entire repositories. Second, rather than merely capturing the final code or the associated chain-of-thought, we reconstruct the entire agentic process of developing a repository. This involves synthesizing a sequence of actions, tool interactions, and evolving states, thereby providing a more comprehensive and realistic representation of the software development lifecycle.

\section{Approach}
\begin{figure*}
    \centering
    \includegraphics[width=\linewidth]{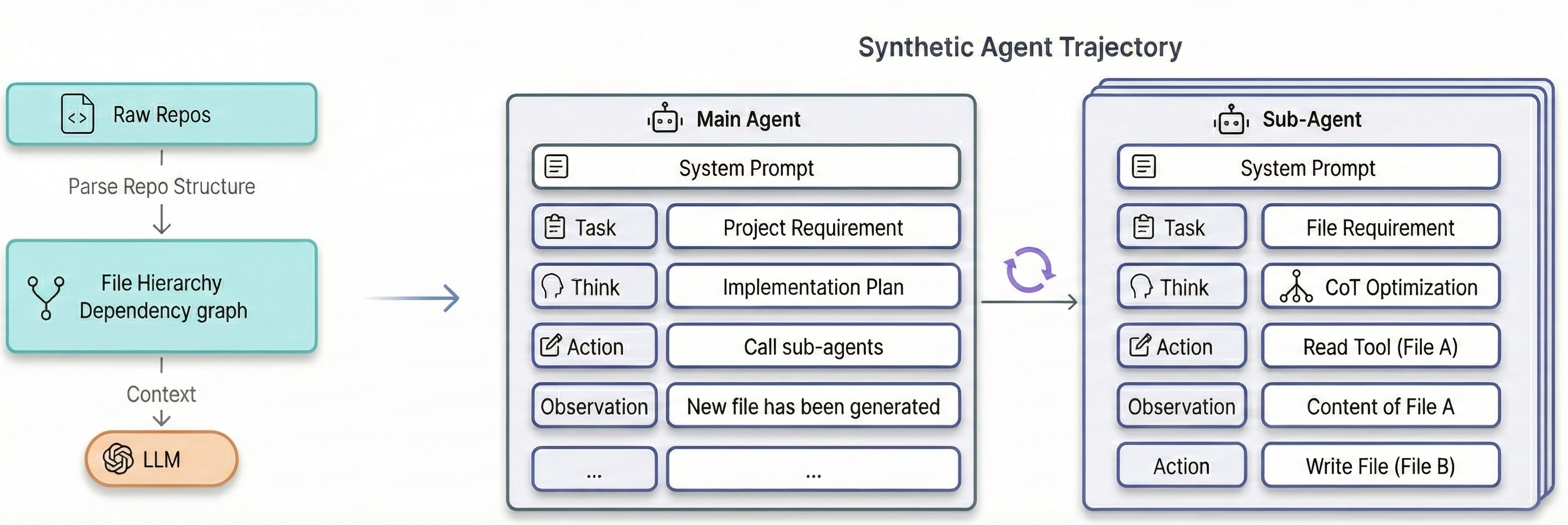}
    \caption{The pipeline of synthetic agent trajectory curation.}
    \label{fig:workflow}
\end{figure*}

Our goal is to create a high-quality, structured dataset of agentic trajectories from existing code repositories for LLM pretraining. Our method consists of two main stages: (1) Multi-Agent Trajectory Curation, where we simulate a developer workflow to reverse-engineer an agentic trajectory from a complete repository, and (2) LongCoT Optimization, where we refine the reasoning within these trajectories using a search-based algorithm. Figure \ref{fig:workflow} provides an overview of our entire pipeline.

The primary objective of this framework is not agent training, but the curation of a high-fidelity pre-training corpus. By simulating the development process, we provide the model with a reasoning-dense supervision signal characterized by long-horizon context dependencies that go far beyond surface-level code patterns.

\subsection{Multi-Agent Trajectory Curation}

\begin{table*}[t]
\centering
\resizebox{\textwidth}{!}{%
\begin{tabularx}{1.2\textwidth}{cllX}
\toprule
\textbf{Step} & \textbf{Role} & \textbf{Component} & \textbf{Content} \\
\midrule
0 & User & \textbf{Task} & Create a simple calculator project in Python. It should include a file for arithmetic operations and a main script to run a demonstration. \\
\addlinespace
1 & Main Agent & \textcolor{blue!70!black}{\textbf{Think}} & The project requires a calculator. It needs a file for core logic, \code{operations.py}, and a main file to run it, \code{main.py}. The latter depends on the former, so I must create \code{operations.py} first. The plan is: 1. Create \code{operations.py}. 2. Create \code{main.py}. \\
\addlinespace
2 & Main Agent & \textcolor{red!70!black}{\textbf{Action}} & \code{CallSubAgent(task="Create 'operations.py'...")} \\
\addlinespace
... & ... & ... & ... \\
\addlinespace
5 & Main Agent & \textcolor{green!50!black}{\textbf{Observation}} & Sub-agent finished task for \code{operations.py}. The file has been generated. \\
\addlinespace
6 & Main Agent & \textcolor{blue!70!black}{\textbf{Think}} & Now proceeding to the next step in my plan. \\
\addlinespace
7 & Main Agent & \textcolor{red!70!black}{\textbf{Action}} & \code{CallSubAgent(task="Create 'main.py' that uses the add function...")} \\
\addlinespace
8 & Sub-Agent & \textcolor{blue!70!black}{\textbf{Think}} & I need to create \code{main.py}. This file must import the \code{add} function from \code{operations.py}. To ensure correctness, I should first read the content of \code{operations.py}. \\
\addlinespace
9 & Sub-Agent & \textcolor{red!70!black}{\textbf{Action}} & \code{ReadTool(file="operations.py")} \\
\addlinespace
10 & Sub-Agent & \textcolor{green!50!black}{\textbf{Observation}} & \code{def add(a, b):\string\n    return a + b} \\
\addlinespace
11 & Sub-Agent & \textcolor{blue!70!black}{\textbf{Think}} & Okay, I have the content of \code{operations.py}. Now I can write \code{main.py}. I will import \code{add}, call it with 2 and 3, and print the output. \\
\addlinespace
12 & Sub-Agent & \textcolor{red!70!black}{\textbf{Action}} & \code{WriteTool(file="main.py", content="from operations import add...")} \\
\addlinespace
13 & Sub-Agent & \textcolor{green!50!black}{\textbf{Observation}} & Successfully wrote 89 bytes to \code{main.py}. \\
\addlinespace
14 & Main Agent & \textcolor{blue!70!black}{\textbf{Think}} & Sub-agent finished task for \code{main.py}. All files in the plan have been generated. The project is complete. \\
\bottomrule
\end{tabularx}%
}
\caption{An illustrative example of a synthesized agent trajectory for creating a tiny project. The process starts with a user instruction, followed by the interaction between the Main Agent and Sub-Agents.}
\label{tab:trajectory_example}
\end{table*}

We design a multi-agent workflow that mirrors a human software development process. Table \ref{tab:trajectory_example} shows an illustrative example of such a synthesized trajectory. Instead of building a live agent framework, we prompt a powerful LLM to simulate the entire workflow and generating the corresponding trajectory data. The simulation unfolds as follows:

\paragraph{Main Agent: Project Planning}
The process begins with a Main Agent. Its responsibilities are high-level planning and coordination. Given the entire code repository as context, the Main Agent is prompted to:
\begin{itemize}
    \item Generate Project Requirements: Synthesize a high-level description of the project's purpose and functionality, as if it were a task brief.
    \item Formulate an Implementation Plan: Decompose the project into a logical sequence of file creation steps. This plan outlines which files should be created and in what order, establishing a dependency-aware development path.
\end{itemize}

For each file in the implementation plan, the Main Agent then invokes a specialized Sub-Agent to handle the implementation.

\paragraph{Sub-Agent: File Implementation}
A Sub-Agent is responsible for generating the code for a single file. This process is also broken down into thought and action steps:
\begin{itemize}
    \item Plan File Implementation: The Sub-Agent first outlines a plan for the specific file's structure and logic.
    \item Information Gathering (\code{Read} Tool): Before writing code, the Sub-Agent may need to understand the context of other parts of the repository. It simulates this by calling a \code{Read} tool to access the content of other, already ``implemented'' files.
    \item Code Generation (\code{Write}  Tool): Finally, the Sub-Agent calls a \code{Write} tool, providing the full code content for the current file.
\end{itemize}

This entire sequence of thoughts, tool calls (\code{Read}, \code{Write}), and tool responses constitutes a single, coherent agentic trajectory (see Steps 8-13 in Table \ref{tab:trajectory_example} for a detailed instance).

\paragraph{Grounding the Simulation with Extracted Information}
A purely LLM-simulated trajectory is prone to noise and hallucinations. To enhance the fidelity and accuracy of our synthetic data, we ground the simulation by injecting ground-truth information extracted directly from the source repository. This serves two purposes: guiding the LLM's generation and replacing noisy outputs with factual data.

We extract the following ground-truth information:
\begin{itemize}
    \item File Structure Tree: A complete directory and file layout of the repository. This is provided to the LLM to simulate the implementation plan of Main Agent.
    \item Inter-File Dependency Graph: We analyze import statements to build a graph representing how files depend on one another. This is important for the LLM to simulate the tool call and tool response of \code{Read} Tool. 
    \item  Intra-File Structure: For each file, we parse its Abstract Syntax Tree (AST) to extract key structural elements like class and function definitions. This information is provided to the LLM to simulate the Sub-Agent trajectory.
\end{itemize}

Furthermore, we use this ground-truth data to correct parts of the simulated trajectory. For example:
\begin{itemize}
    \item The response to a \code{Read} tool call is replaced with the actual content of the file from the repository.
    \item The final output of the \code{Write} tool call is replaced with the ground-truth code of the file.

\end{itemize}
This grounding process ensures that while the reasoning is generated by the LLM, the actions and outcomes are anchored to reality.

\subsection{CoT Optimization via Search}
\label{sec:optimize-agent-memory}

The initial trajectory curation stage leverages an LLM's ability to simulate agentic behavior. However, the generated CoT reasoning ($z$) may not be optimal for generating the target code ($x$). An ideal thought process should make the subsequent code generation step as simple as possible. Formally, we aim to find a reasoning path $z^*$ that maximizes the conditional log-probability of the code:

\begin{equation}
    z^* = \arg\max_{z} \log p(x | z)
\end{equation}

While this objective could be optimized using RL (with $\log p(x|z)$ as the reward), RL training is often complex, expensive, and unstable. We therefore opt for a simpler yet effective inference-time search strategy.

Following \citet{reer}, we decompose the CoT into steps $(z_1, \dots, z_n)$ and optimize each $z_i$:

\begin{enumerate}
    \item Sample: We prompt an LLM to generate a set of $k$ alternative ``refinements'' for the thought step $z_i$.
    \item Evaluate: For each candidate $z_{cand} = (z_1, \dots, z'_i, \dots, z_n)$, measure the Perplexity (PPL) of the ground-truth code $x$: $\text{PPL}(x | z_{cand})$.
    \item Update: If the best refinement $z'^*_{i}$ results in a lower perplexity than the original step $z_i$, we permanently update the CoT with this new, improved step.
\end{enumerate}

This iterative refinement ensures the reasoning path is causally structured and directly facilitates correct code generation.

\subsection{Continue Pretraining on Synthetic Agent Trajectories}
We utilize the synthesized agent trajectories for continual pre-training rather than SFT or post-training. This choice is motivated by the inherent nature of our synthetic data. The trajectories inevitably contain noise and biases stemming from the LLM's potential hallucinations and our agent workflow. Continuous pre-training, which typically involves larger and more diverse datasets than SFT, is inherently more robust to such imperfections.

\paragraph{Trajectory Flattening} 
To prepare the data, we transform the hierarchical multi-agent interaction into a single sequential document. When the Main Agent calls a Sub-Agent, we recursively inject that Sub-Agent's entire trajectory (thoughts, tool calls, and observations) directly into the call point. This creates a monolithic, chronological sequence that mirrors the complete development lifecycle of the repository, structurally similar to the example shown in Table \ref{tab:trajectory_example}.

\paragraph{Targeted Loss Masking} 
To ensure the model learns the causal link between reasoning and action rather than memorizing feedback, we mask the tokens corresponding to \textbf{Observations} (tool responses). The model is thus trained exclusively to predict \code{Think} and \code{Action} tokens, forcing it to internalize the logic of the development process.
\section{Experiments}
\subsection{Experiment Setup}
\paragraph{Data Generation:} We curate approximately 300k GitHub repositories by filtering too short and long repositories. Using Qwen3-30B-A3B-Instruct-2507 \citep{qwen3technicalreport}, we generate 4B tokens of synthetic agent trajectories. For CoT optimization, we generate two candidates for each CoT step and iterate the search-and-replace process for 3 rounds.

\paragraph{Training Configuration:} We continually pre-train Llama3-8B-Instruct \citep{llama3} for 20B tokens with a 64k context window, following \citet{prolong}. To ensure a fair comparison, all models share a 70\% general-domain and 30\% repository-related data mixture. Within the 30\% repository slot, 18\% is fixed (Prolong Repos), while the remaining 12\% is allocated to our experimental data variants.

\paragraph{Baselines and Model Variants:} We compare the official Prolong baseline against three internal variants, differing only in the 12\% experimental data slot:
\begin{itemize}
\item Raw-Repos: 12\% slot filled with raw source code from our 300k repos.
\item Repo2Agent: 12\% slot filled with unoptimized synthetic trajectories.
\item Repo2Agent-Search: 12\% slot filled with search-optimized trajectories.
\end{itemize}

Our primary comparison focuses on \textit{Raw-Repos}, \textit{Repo2Agent}, and \textit{Repo2Agent-Search} to isolate the impact of converting code into agentic trajectories.

\paragraph{Evaluation Benchmarks}
We assess the models across four key capabilities. The selection of these benchmarks is directly motivated by the long-context, reasoning-intensive, and code-centric nature of our reconstructed data.

\begin{itemize}
\item Long-Context Understanding: As our reconstruction unfolds repositories into massive sequential traces, it introduces long-range causal dependencies. we evaluate this via Ruler \citep{hsieh2024ruler} and Helmet \citep{yen2025helmet} to test information retrieval across extended horizons.
\item Coding: Given our code-domain focus, we use LongCodeBench \citep{rando2025longcodebench} and HumanEval \citep{humaneval} to verify if observing the ``process'' of code creation enables better synthesis than memorizing static files.
\item Reasoning: A central feature of our data is the search-optimized CoT. We evaluate the transferability of this structured logic to general domains using BBH \citep{bbh-hard}, AGIEval \citep{zhong2024agieval}, GSM-8k \citep{gsm8k}, MATH \citep{math}, and MMLU-Pro \citep{wang2024mmlu}.
\item Software-engineering capability: We use APTBench \citep{qin2025aptbench}, which is specifically designed to assess the foundational agentic capabilities of pre-trained models (without post-training) on SWE-Bench and Deep-Research, to measure the inherent potential instilled by our trajectories.
\end{itemize}

\begin{table*}[t]
\centering
\begin{tabular}{@{}llcccc@{}}
\toprule
Benchmark & Context & Prolong & Raw-Repo & Repo2Agent & Repo2Agent-Search \\ \midrule
\multirow{3}{*}{Ruler} 
    & 16,384 & 83.61 & 86.90 & \textbf{87.50} & 87.10 \\
    & 32,768 & 81.77 & 83.20 & 84.00 & \textbf{84.40} \\
    & 65,536 & 57.10 & 61.00 & 58.10 & \textbf{61.80} \\ 
\midrule
\multirow{3}{*}{Helmet} 
    & 16,384 & 60.17 & 60.41 & 61.56 & \textbf{61.99} \\
    & 32,768 & 61.57 & 60.98 & 62.03 & \textbf{62.65} \\
    & 65,536 & \textbf{58.10} & 57.13 & 57.32 & 57.84 \\ 
\bottomrule
\end{tabular}
\caption{Summary of Long-Context Understanding performance (Average Scores). Detailed sub-task results are provided in Appendix \ref{app:detailed-long-bench}.}
\label{tab:long-context-summary}
\end{table*}
\subsection{Main Results}
\subsubsection{Long-Context Understanding}
We evaluate the long-context capabilities of our models using two comprehensive benchmarks: Ruler and Helmet. Across both benchmarks, our primary observation is that training on structured agent trajectories (Repo2Agent variants) consistently yields superior performance compared to training on flattened code (Raw-Repos). This confirms that reconstructing the process of code generation provides a denser, more instructive signal for long-context modeling than static code files alone. Furthermore, our optimized model, \textit{Repo2Agent-Search}, frequently surpasses the strong external baseline (\textit{Prolong}), particularly in tasks requiring complex information retrieval.

We present the average scores of long-context understanding in Table \ref{tab:long-context-summary}. For a more granular breakdown of performance across all sub-tasks in Ruler and Helmet, please refer to Appendix \ref{app:detailed-long-bench} (Tables \ref{tab:ruler} and \ref{tab:helmet}).

\paragraph{Performance on Ruler}
As shown in Table \ref{tab:long-context-summary}, the models trained on our synthetic data (Repo2Agent \& Repo2Agent-Search) consistently outperform the internal \textit{Raw-Repo} baseline across all tested context lengths.

At shorter context lengths (16k and 32k), \textit{Repo2Agent} and \textit{Repo2Agent-Search} maintain a steady lead over raw code pre-training. The advantage of agentic synthetic data becomes most evident at the 64k window size. While the official \textit{Prolong} baseline and the \textit{Raw-Repo} ablation show significant degradation, \textit{Repo2Agent-Search} achieves the highest robustness with an average score of 61.80. This suggests that learning from a structured, step-by-step construction process helps the model maintain information integrity even when the context is heavily populated.

\paragraph{Performance on Helmet}
The results on the Helmet benchmark further reinforce the superiority of the reconstruction paradigm.

At 16k and 32k context lengths, \textit{Repo2Agent-Search} achieves peak performance, reaching an average of 62.65 at 32k. This represents a significant improvement over the \textit{Raw-Repo} baseline (60.98), indicating that the search-optimized reasoning steps provide a cleaner and more effective supervision signal for long-range retrieval and reasoning than flattened code files. At the maximum length of 64k, while the \textit{Prolong} baseline remains highly competitive, our \textit{Repo2Agent-Search} continues to outperform the primary \textit{Raw-Repo} ablation. This confirms that even when considering the holistic performance across diverse long-context tasks, converting static repositories into dynamic histories is a more potent data strategy than standard code pre-training.

\begin{table*}[t]
    \centering
    \begin{tabular}{llccc}
        \toprule
        Benchmark       &Prolong& Raw-Repos & Repo2Agent & Repo2Agent-Search \\
        \midrule
        AGI-Eval          &\textbf{36.91}& 35.78    & 36.32    & 36.85   \\
        BBH             &66.69& 66.27    & 66.00    & \textbf{67.03}   \\
        GSM-8k           &59.67& \textbf{61.94}    & \textbf{61.94}    & 60.96   \\
        MATH            &1.64&  2.18    & 3.72    & \textbf{3.76}   \\
        Human-Eval      &16.46& 34.76    & 36.59    & \textbf{37.20}   \\
        LongCodeBench-32k  &29.38& 34.16 & 34.51 & \textbf{36.46} \\
        LongCodeBench-64k  & 30.52 & 27.37 & \textbf{31.05} & 30.26 \\
        \bottomrule
    \end{tabular}
    \caption{Results on Reasoning or Coding Benchmarks, including AGI-Eval, BBH, GSM-8k, MATH, Human-Eval and LongCodeBench.}
    \label{tab:code-and-reasoning}
\end{table*}
\subsubsection{Coding and Reasoning}
We further evaluate whether agentic pre-training benefits fundamental coding and general reasoning (Table \ref{tab:code-and-reasoning}).

\paragraph{Coding Capabilities} Our reconstruction paradigm shows a clear advantage in code generation. On HumanEval, \textit{Repo2Agent-Search} scores 37.20, outperforming the \textit{Raw-Repos} baseline (34.76). This confirms that learning the ``process'' of creation—incorporating planning and refinement—is superior to memorizing static code. This edge extends to long-horizon tasks; \textit{Repo2Agent-Search} leads on LongCodeBench-32k (36.46). 

\paragraph{Reasoning Transfer} Despite the lack of math-specific tuning, our method induces positive transfer to general reasoning. On MATH, although absolute scores are low across all models—reflecting the inherent limitations of the Llama-3-8B in complex mathematics—\textit{Repo2Agent-Search} still yields the best results. Furthermore, on BBH and AGI-Eval, our models match or slightly exceed the baselines. These results demonstrate that the structured logic within agentic trajectories provides a higher-quality supervision signal than raw code, enhancing specialized skills without compromising general intelligence.

\begin{table*}[t]
  \centering
  \sisetup{table-format=2.2, table-space-text-post={\%}, group-minimum-digits=2}
  \begin{tabular}{llccc} 
    \toprule
    \textbf{Category} & \textbf{Sub-task} & \textbf{Raw-Repos} & \textbf{Repo2Agent} & \textbf{Repo2Agent-Search} \\
    \midrule
    \multirow{6}{*}{DeepResearch} & Openend-Citation & 11.20 & 10.94 & \textbf{11.49} \\
                                   & Openend-Plan     & \textbf{16.11} & 13.42 & 10.40 \\
                                   & Openend-Quality  & 21.99 & 24.74 & \textbf{26.20} \\
                                   & Plan             & 47.09 & \textbf{49.54} & 47.81 \\
                                   & Summ-Ans         & 43.12 & \textbf{45.30} & 44.40 \\
                                   & \textit{Average} & 29.21 & \textbf{30.49} & 30.02 \\
    \midrule
    \multirow{4}{*}{Env-Setup}    & Action           & 20.39 & \textbf{22.05} & 21.13 \\
                                   & Error            & 22.45 & 23.13 & \textbf{24.49} \\
                                   & Plan             & 18.99 & 17.85 & \textbf{19.22} \\
                                   & \textit{Average} & 20.61 & 21.01 & \textbf{21.61} \\
    \midrule
    \multirow{6}{*}{Issue-Fix}    & Fix-Patch        & 26.72 & \textbf{28.02} & 25.43 \\
                                   & Locate           & \textbf{24.03} & 23.67 & \textbf{24.03} \\
                                   & Plan             & 37.04 & \textbf{40.74} & 38.68 \\
                                   & Test-Patch       & 26.60 & \textbf{27.08} & 26.60 \\
                                   & Tool-Call        & 54.23 & \textbf{54.69} & 54.28 \\
                                   & \textit{Average} & 33.72 & \textbf{34.84} & 33.80 \\
    \midrule
    \multicolumn{2}{l}{\textbf{Overall Average}} & 29.02 & \textbf{30.10} & 29.65 \\
    \bottomrule
  \end{tabular}
    \caption{Results on APTBench (Merged En/Zh Sub-tasks)}
      \label{tab:detailed_performance_compressed}
\end{table*}
\subsubsection{Software-Engineering Capability}
We use APTBench to evaluate the foundational argentic potential instilled by our pre-training. By deconstructing complex trajectories into atomic skills (e.g., planning, fix-patch, tool selection, test-patch, ...), APTBench measures a model's inherent aptitude without the confounding effects of post-training.

\textit{Repo2Agent} excels in planning-centric categories like Issue-Fix (34.84\%). This suggests that natural, unrefined CoT provides a more generalizable signal for holistic workflows. \textit{Repo2Agent-Search} leads in Env-Setup (21.61\%), particularly in the Error diagnosis sub-task (24.49\%). This indicates that search-refined reasoning, being logically more rigorous, is more effective for teaching meticulous, low-level implementation and debugging logic.

In summary, pre-training on synthetic trajectories (Repo2Agent) significantly fosters innate agentic capabilities compared to raw code, with the choice of optimization (Search) offering a tunable balance between broad planning and logical precision.

\begin{figure*}[t]
    \centering
    \begin{subfigure}[b]{0.58\textwidth}
        \centering
        \includegraphics[width=\linewidth, height=6cm]{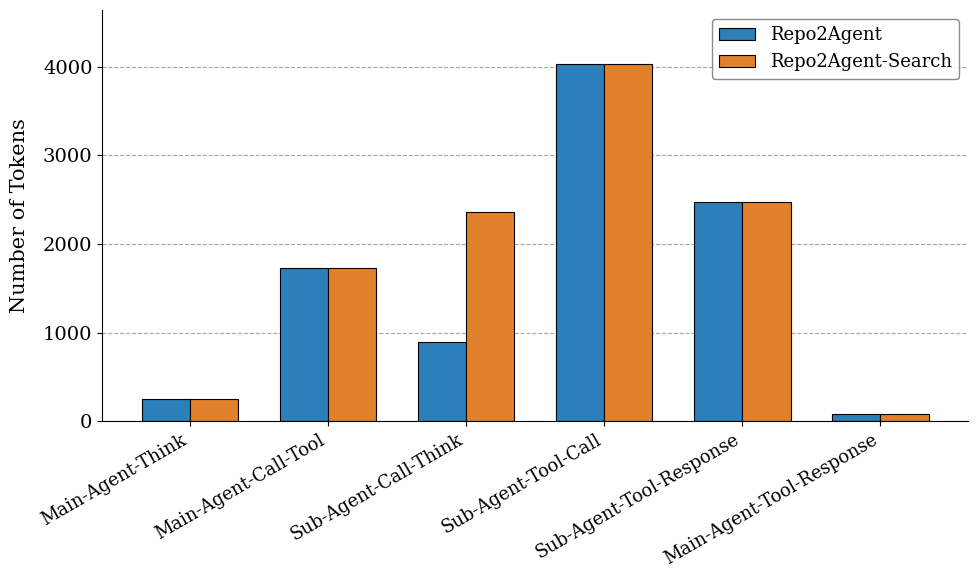} 
        \caption{Composition of Agent Trajectories}
        \label{fig:agent-composition}
    \end{subfigure}
    \hfill
    \begin{subfigure}[b]{0.4\textwidth}
        \centering
        \includegraphics[width=\linewidth, height=5cm]{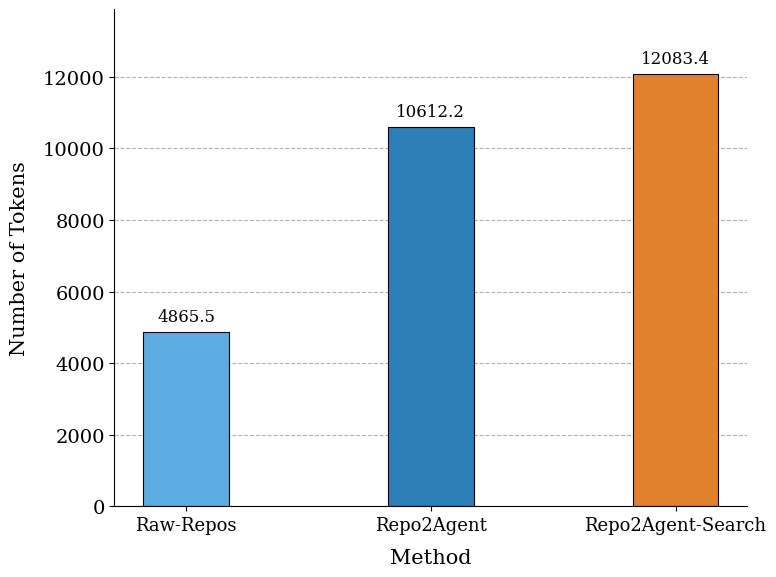} 
        \caption{Trajectory Length}
        \label{fig:length-compare}
    \end{subfigure}
    \caption{(a): Token distribution on thinking, too-call and too-response of main-agent and sub-agent. (b): Average number of tokens for each repo.}
    \label{fig:combined}
\end{figure*}
\begin{figure*}[t]
    \centering
    \begin{subfigure}[b]{0.45\linewidth}
        \centering
        \includegraphics[width=\linewidth]{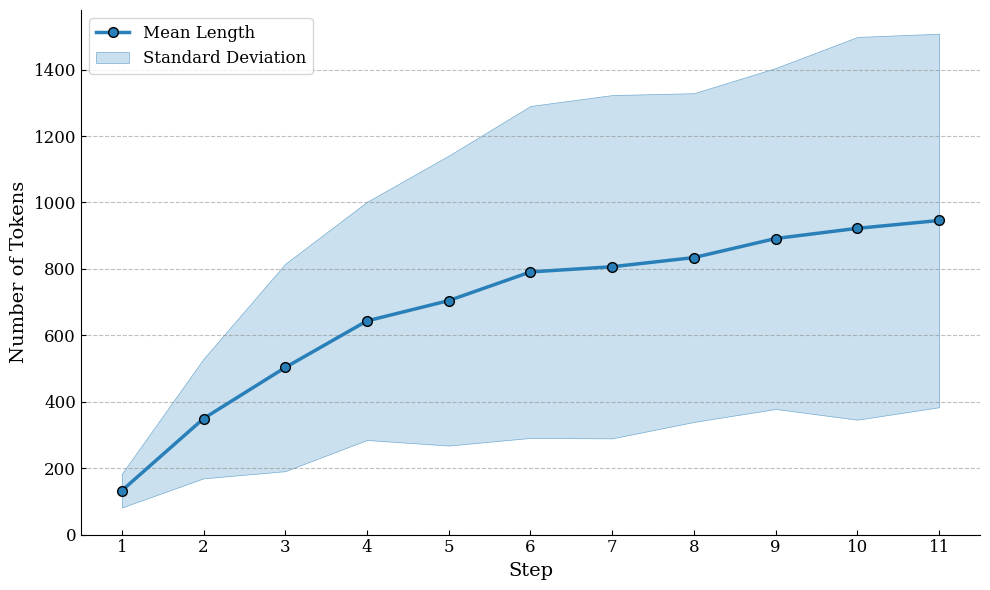}  
        \caption{CoT Length}  
        \label{fig:longer-cot}  
    \end{subfigure}
    \hfill  
    \begin{subfigure}[b]{0.45\linewidth}
        \centering
        \includegraphics[width=\linewidth]{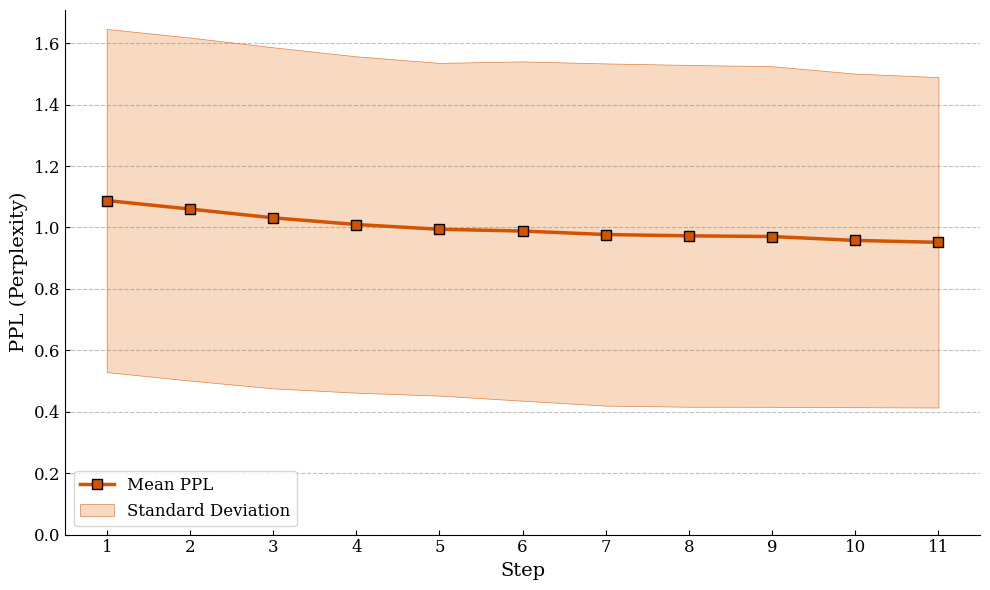}
        \caption{PPL of the codes }  
        \label{fig:lower-ppl}  
    \end{subfigure}
    
    \caption{(a): The CoT increases with more CoT-optimization iterations. (b): PPL of the code to be generated decreases with more iterations.}
    \label{fig:combined-figures}  
\end{figure*}
\section{Analysis on Synthetic Data}
\subsection{Token Distribution}
We analyze the structural composition and length of our synthetic trajectories to evaluate the impact of agentic reconstruction and search-based optimization (Figure \ref{fig:combined}).

\paragraph{Composition and Reasoning Expansion} As shown in Figure \ref{fig:agent-composition}, tokens are primarily concentrated in sub-agent activities (Tool-Calls and Responses), reflecting the detailed implementation process. Crucially, our search optimization significantly deepens the reasoning trace: Sub-Agent-Call-Think tokens more than double from ~900 in \textit{Repo2Agent} to ~2,300 in \textit{Repo2Agent-Search}. This validates that the search process doesn't merely refine thoughts but substantially elaborates on the logical steps required for implementation.

\paragraph{Information Expansion} Figure \ref{fig:length-compare} highlights how our paradigm decompresses static code into explicit narratives.
Transforming raw code (avg. 4,865.5 tokens) into an agentic trajectory (\textit{Repo2Agent-Search}, avg. 12,083.4 tokens) significantly increases the per-repository token count by making latent planning and execution steps explicit.

Importantly, despite the increased sample length, all model variants are trained on a fixed budget of 12\% of 20B total tokens. This ensures a fair comparison: the performance gains are driven by the structural quality and informational density of the trajectories, rather than an increase in the total volume of training data.

\subsection{Impact of CoT Optimization}
We evaluate the relationship between optimization iterations, CoT length, and target code perplexity (PPL) using 100 sample trajectories over 10 iterations (Figure \ref{fig:combined-figures}).

As shown in Figure \ref{fig:longer-cot}, the average CoT length correlates positively with the number of iterations, confirming that our search-based method actively elaborates on the initial reasoning to produce more explicit thought processes. Crucially, this elaboration directly improves reasoning quality: Figure \ref{fig:lower-ppl} illustrates a steady decrease in code PPL as iterations increase. This inverse relationship supports our hypothesis that more detailed reasoning provides a more informative and predictive context, thereby simplifying the subsequent code generation task.

\section{Conclusion}
In this work, we addressed the limitations of training on static software artifacts by proposing a novel paradigm of \textbf{understanding via reconstruction}. By reverse-engineering latent agentic trajectories through grounded multi-agent simulation and refining the reasoning via search-based optimization, we transformed static repositories into dynamic, causally rich training data. Our experiments with Llama-3-8B demonstrate that learning from these reconstructed processes significantly enhances coding, reasoning, and agentic capabilities. 

\clearpage
    
\bibliographystyle{plainnat}
\bibliography{main}

\begin{thebibliography}{39}
\providecommand{\natexlab}[1]{#1}
\providecommand{\url}[1]{\texttt{#1}}
\expandafter\ifx\csname urlstyle\endcsname\relax
  \providecommand{\doi}[1]{doi: #1}\else
  \providecommand{\doi}{doi: \begingroup \urlstyle{rm}\Url}\fi

\bibitem[Chaudhary(2023)]{codealpaca}
Sahil Chaudhary.
\newblock Code alpaca: An instruction-following llama model for code generation.
\newblock \url{https://github.com/sahil280114/codealpaca}, 2023.

\bibitem[Chen(2021)]{humaneval}
Mark Chen.
\newblock Evaluating large language models trained on code.
\newblock \emph{arXiv preprint arXiv:2107.03374}, 2021.

\bibitem[Chen et~al.(2025)Chen, Zhao, Zhang, Liu, Qi, Wu, Kalluri, Cao, Xiong, Tong, et~al.]{scale-agent-traj}
Zhaorun Chen, Zhuokai Zhao, Kai Zhang, Bo~Liu, Qi~Qi, Yifan Wu, Tarun Kalluri, Sara Cao, Yuanhao Xiong, Haibo Tong, et~al.
\newblock Scaling agent learning via experience synthesis.
\newblock \emph{arXiv preprint arXiv:2511.03773}, 2025.

\bibitem[Cobbe et~al.(2021)Cobbe, Kosaraju, Bavarian, Chen, Jun, Kaiser, Plappert, Tworek, Hilton, Nakano, et~al.]{gsm8k}
Karl Cobbe, Vineet Kosaraju, Mohammad Bavarian, Mark Chen, Heewoo Jun, Lukasz Kaiser, Matthias Plappert, Jerry Tworek, Jacob Hilton, Reiichiro Nakano, et~al.
\newblock Training verifiers to solve math word problems.
\newblock \emph{arXiv preprint arXiv:2110.14168}, 2021.

\bibitem[Dubey et~al.(2024)Dubey, Jauhri, Pandey, Kadian, Al-Dahle, Letman, Mathur, Schelten, Yang, Fan, et~al.]{llama3}
Abhimanyu Dubey, Abhinav Jauhri, Abhinav Pandey, Abhishek Kadian, Ahmad Al-Dahle, Aiesha Letman, Akhil Mathur, Alan Schelten, Amy Yang, Angela Fan, et~al.
\newblock The llama 3 herd of models.
\newblock \emph{arXiv e-prints}, pages arXiv--2407, 2024.

\bibitem[Fang et~al.(2025)Fang, Cai, Li, Wu, Li, Yin, Wang, Wang, Su, Zhang, et~al.]{qwen-scale-env}
Runnan Fang, Shihao Cai, Baixuan Li, Jialong Wu, Guangyu Li, Wenbiao Yin, Xinyu Wang, Xiaobin Wang, Liangcai Su, Zhen Zhang, et~al.
\newblock Towards general agentic intelligence via environment scaling.
\newblock \emph{arXiv preprint arXiv:2509.13311}, 2025.

\bibitem[Floridi and Chiriatti(2020)]{floridi2020gpt}
Luciano Floridi and Massimo Chiriatti.
\newblock Gpt-3: Its nature, scope, limits, and consequences.
\newblock \emph{Minds and machines}, 30\penalty0 (4):\penalty0 681--694, 2020.

\bibitem[Gao et~al.(2025)Gao, Wettig, Yen, and Chen]{prolong}
Tianyu Gao, Alexander Wettig, Howard Yen, and Danqi Chen.
\newblock How to train long-context language models (effectively).
\newblock In \emph{Proceedings of the 63rd Annual Meeting of the Association for Computational Linguistics (Volume 1: Long Papers)}, pages 7376--7399, 2025.

\bibitem[Guo et~al.(2024)Guo, Zhu, Yang, Xie, Dong, Zhang, Chen, Bi, Wu, Li, Luo, Xiong, and Liang]{deepseek-coder}
Daya Guo, Qihao Zhu, Dejian Yang, Zhenda Xie, Kai Dong, Wentao Zhang, Guanting Chen, Xiao Bi, Y.~Wu, Y.K. Li, Fuli Luo, Yingfei Xiong, and Wenfeng Liang.
\newblock Deepseek-coder: When the large language model meets programming -- the rise of code intelligence, 2024.
\newblock URL \url{https://arxiv.org/abs/2401.14196}.

\bibitem[Hendrycks et~al.(2021)Hendrycks, Burns, Kadavath, Arora, Basart, Tang, Song, and Steinhardt]{math}
Dan Hendrycks, Collin Burns, Saurav Kadavath, Akul Arora, Steven Basart, Eric Tang, Dawn Song, and Jacob Steinhardt.
\newblock Measuring mathematical problem solving with the math dataset.
\newblock \emph{arXiv preprint arXiv:2103.03874}, 2021.

\bibitem[Hsieh et~al.(2024)Hsieh, Sun, Kriman, Acharya, Rekesh, Jia, Zhang, and Ginsburg]{hsieh2024ruler}
Cheng-Ping Hsieh, Simeng Sun, Samuel Kriman, Shantanu Acharya, Dima Rekesh, Fei Jia, Yang Zhang, and Boris Ginsburg.
\newblock Ruler: What's the real context size of your long-context language models?
\newblock \emph{arXiv preprint arXiv:2404.06654}, 2024.

\bibitem[Hui et~al.(2024)Hui, Yang, Cui, Yang, Liu, Zhang, Liu, Zhang, Yu, Dang, et~al.]{hui2024qwen2}
Binyuan Hui, Jian Yang, Zeyu Cui, Jiaxi Yang, Dayiheng Liu, Lei Zhang, Tianyu Liu, Jiajun Zhang, Bowen Yu, Kai Dang, et~al.
\newblock Qwen2. 5-coder technical report.
\newblock \emph{arXiv preprint arXiv:2409.12186}, 2024.

\bibitem[Jimenez et~al.(2023)Jimenez, Yang, Wettig, Yao, Pei, Press, and Narasimhan]{jimenez2023swe}
Carlos~E Jimenez, John Yang, Alexander Wettig, Shunyu Yao, Kexin Pei, Ofir Press, and Karthik Narasimhan.
\newblock Swe-bench: Can language models resolve real-world github issues?
\newblock \emph{arXiv preprint arXiv:2310.06770}, 2023.

\bibitem[Li et~al.(2025)Li, Inan, Yue, Chen, Wutschitz, Kulkarni, Poovendran, Sim, and Rajmohan]{sim-rl}
Yuetai Li, Huseyin~A Inan, Xiang Yue, Wei-Ning Chen, Lukas Wutschitz, Janardhan Kulkarni, Radha Poovendran, Robert Sim, and Saravan Rajmohan.
\newblock Simulating environments with reasoning models for agent training.
\newblock \emph{arXiv preprint arXiv:2511.01824}, 2025.

\bibitem[Luo et~al.(2023)Luo, Xu, Zhao, Sun, Geng, Hu, Tao, Ma, Lin, and Jiang]{luo2023wizardcoder}
Ziyang Luo, Can Xu, Pu~Zhao, Qingfeng Sun, Xiubo Geng, Wenxiang Hu, Chongyang Tao, Jing Ma, Qingwei Lin, and Daxin Jiang.
\newblock Wizardcoder: Empowering code large language models with evol-instruct.
\newblock \emph{arXiv preprint arXiv:2306.08568}, 2023.

\bibitem[Ouyang et~al.(2022)Ouyang, Wu, Jiang, Almeida, Wainwright, Mishkin, Zhang, Agarwal, Slama, Ray, et~al.]{instruct-gpt}
Long Ouyang, Jeffrey Wu, Xu~Jiang, Diogo Almeida, Carroll Wainwright, Pamela Mishkin, Chong Zhang, Sandhini Agarwal, Katarina Slama, Alex Ray, et~al.
\newblock Training language models to follow instructions with human feedback.
\newblock \emph{Advances in neural information processing systems}, 35:\penalty0 27730--27744, 2022.

\bibitem[Pahuja et~al.(2025)Pahuja, Lu, Rosset, Gou, Mitra, Whitehead, Su, and Hassan]{scale-explorer}
Vardaan Pahuja, Yadong Lu, Corby Rosset, Boyu Gou, Arindam Mitra, Spencer Whitehead, Yu~Su, and Ahmed Hassan.
\newblock Explorer: Scaling exploration-driven web trajectory synthesis for multimodal web agents.
\newblock In \emph{Findings of the Association for Computational Linguistics: ACL 2025}, pages 6300--6323, 2025.

\bibitem[Pang et~al.(2025)Pang, Dong, Xu, Savarese, Zhou, and Xiong]{bolt}
Bo~Pang, Hanze Dong, Jiacheng Xu, Silvio Savarese, Yingbo Zhou, and Caiming Xiong.
\newblock Bolt: Bootstrap long chain-of-thought in language models without distillation, 2025.
\newblock URL \url{https://arxiv.org/abs/2502.03860}.

\bibitem[Pham et~al.(2025)Pham, Phan, Phan, Chi, Nguyen, and Bui]{pham2025swe}
Minh~VT Pham, Huy~N Phan, Hoang~N Phan, Cuong~Le Chi, Tien~N Nguyen, and Nghi~DQ Bui.
\newblock Swe-synth: Synthesizing verifiable bug-fix data to enable large language models in resolving real-world bugs.
\newblock \emph{arXiv preprint arXiv:2504.14757}, 2025.

\bibitem[Qin et~al.(2025)Qin, Xi, Huang, Rui, Yin, Liu, Yu, Zhang, and Sun]{qin2025aptbench}
Jiarui Qin, Yunjia Xi, Junjie Huang, Renting Rui, Di~Yin, Weiwen Liu, Yong Yu, Weinan Zhang, and Xing Sun.
\newblock Aptbench: Benchmarking agentic potential of base llms during pre-training.
\newblock \emph{arXiv preprint arXiv:2510.24397}, 2025.

\bibitem[Qiu et~al.(2024)Qiu, Lu, Zeng, Guo, Geng, Wang, Huang, Wu, and Wang]{qiu2024treebon}
Jiahao Qiu, Yifu Lu, Yifan Zeng, Jiacheng Guo, Jiayi Geng, Huazheng Wang, Kaixuan Huang, Yue Wu, and Mengdi Wang.
\newblock Treebon: Enhancing inference-time alignment with speculative tree-search and best-of-n sampling.
\newblock \emph{arXiv preprint arXiv:2410.16033}, 2024.

\bibitem[Rando et~al.(2025)Rando, Romani, Sampieri, Franco, Yang, Kyuragi, Galasso, and Hashimoto]{rando2025longcodebench}
Stefano Rando, Luca Romani, Alessio Sampieri, Luca Franco, John Yang, Yuta Kyuragi, Fabio Galasso, and Tatsunori Hashimoto.
\newblock Longcodebench: Evaluating coding llms at 1m context windows.
\newblock \emph{arXiv preprint arXiv:2505.07897}, 2025.

\bibitem[Sengupta et~al.(2024)Sengupta, Vashistha, Curtis, Mallipeddi, Mathur, Ross, and Gou]{magv}
Saptarshi Sengupta, Harsh Vashistha, Kristal Curtis, Akshay Mallipeddi, Abhinav Mathur, Joseph Ross, and Liang Gou.
\newblock Mag-v: A multi-agent framework for synthetic data generation and verification.
\newblock \emph{arXiv preprint arXiv:2412.04494}, 2024.

\bibitem[Shao et~al.(2025)Shao, Li, Ma, Li, Song, Cheng, Li, Li, Wang, Guo, et~al.]{shao2025case2code}
Yunfan Shao, Linyang Li, Yichuan Ma, Peiji Li, Demin Song, Qinyuan Cheng, Shimin Li, Xiaonan Li, Pengyu Wang, Qipeng Guo, et~al.
\newblock Case2code: Scalable synthetic data for code generation.
\newblock In \emph{Proceedings of the 31st International Conference on Computational Linguistics}, pages 11056--11069, 2025.

\bibitem[Suzgun et~al.(2022)Suzgun, Scales, Sch{\"a}rli, Gehrmann, Tay, Chung, Chowdhery, Le, Chi, Zhou, , and Wei]{bbh-hard}
Mirac Suzgun, Nathan Scales, Nathanael Sch{\"a}rli, Sebastian Gehrmann, Yi~Tay, Hyung~Won Chung, Aakanksha Chowdhery, Quoc~V Le, Ed~H Chi, Denny Zhou, , and Jason Wei.
\newblock Challenging big-bench tasks and whether chain-of-thought can solve them.
\newblock \emph{arXiv preprint arXiv:2210.09261}, 2022.

\bibitem[Tang et~al.(2023)Tang, Deng, Lin, Han, Liang, Cao, and Sun]{tool-alpaca}
Qiaoyu Tang, Ziliang Deng, Hongyu Lin, Xianpei Han, Qiao Liang, Boxi Cao, and Le~Sun.
\newblock Toolalpaca: Generalized tool learning for language models with 3000 simulated cases.
\newblock \emph{arXiv preprint arXiv:2306.05301}, 2023.

\bibitem[Team et~al.(2025)Team, Bai, Bao, Chen, Chen, Chen, Chen, Chen, Chen, Chen, et~al.]{team2025kimi}
Kimi Team, Yifan Bai, Yiping Bao, Guanduo Chen, Jiahao Chen, Ningxin Chen, Ruijue Chen, Yanru Chen, Yuankun Chen, Yutian Chen, et~al.
\newblock Kimi k2: Open agentic intelligence.
\newblock \emph{arXiv preprint arXiv:2507.20534}, 2025.

\bibitem[Team(2025)]{qwen3technicalreport}
Qwen Team.
\newblock Qwen3 technical report, 2025.
\newblock URL \url{https://arxiv.org/abs/2505.09388}.

\bibitem[Wang et~al.(2025{\natexlab{a}})Wang, Li, Qu, Zhu, Xu, Chu, and Lin]{wang2025code}
Haozhe Wang, Long Li, Chao Qu, Fengming Zhu, Weidi Xu, Wei Chu, and Fangzhen Lin.
\newblock To code or not to code? adaptive tool integration for math language models via expectation-maximization.
\newblock \emph{arXiv preprint arXiv:2502.00691}, 2025{\natexlab{a}}.

\bibitem[Wang et~al.(2025{\natexlab{b}})Wang, Que, Xu, Liu, Zhou, Feng, Zhong, Ye, Yang, Huang, et~al.]{reer}
Haozhe Wang, Haoran Que, Qixin Xu, Minghao Liu, Wangchunshu Zhou, Jiazhan Feng, Wanjun Zhong, Wei Ye, Tong Yang, Wenhao Huang, et~al.
\newblock Reverse-engineered reasoning for open-ended generation.
\newblock \emph{arXiv preprint arXiv:2509.06160}, 2025{\natexlab{b}}.

\bibitem[Wang et~al.(2025{\natexlab{c}})Wang, Xu, Liu, Wu, Lin, and Chen]{wang2025emergent}
Haozhe Wang, Qixin Xu, Che Liu, Junhong Wu, Fangzhen Lin, and Wenhu Chen.
\newblock Emergent hierarchical reasoning in llms through reinforcement learning.
\newblock \emph{arXiv preprint arXiv:2509.03646}, 2025{\natexlab{c}}.

\bibitem[Wang et~al.(2025{\natexlab{d}})Wang, Yang, Huang, Dong, and Wei]{tpt}
Liang Wang, Nan Yang, Shaohan Huang, Li~Dong, and Furu Wei.
\newblock Thinking augmented pre-training, 2025{\natexlab{d}}.
\newblock URL \url{https://arxiv.org/abs/2509.20186}.

\bibitem[Wang et~al.(2023)Wang, Kordi, Mishra, Liu, Smith, Khashabi, and Hajishirzi]{self-instruct}
Yizhong Wang, Yeganeh Kordi, Swaroop Mishra, Alisa Liu, Noah~A Smith, Daniel Khashabi, and Hannaneh Hajishirzi.
\newblock Self-instruct: Aligning language models with self-generated instructions.
\newblock In \emph{Proceedings of the 61st annual meeting of the association for computational linguistics (volume 1: long papers)}, pages 13484--13508, 2023.

\bibitem[Wang et~al.(2024)Wang, Ma, Zhang, Ni, Chandra, Guo, Ren, Arulraj, He, Jiang, et~al.]{wang2024mmlu}
Yubo Wang, Xueguang Ma, Ge~Zhang, Yuansheng Ni, Abhranil Chandra, Shiguang Guo, Weiming Ren, Aaran Arulraj, Xuan He, Ziyan Jiang, et~al.
\newblock Mmlu-pro: A more robust and challenging multi-task language understanding benchmark.
\newblock \emph{Advances in Neural Information Processing Systems}, 37:\penalty0 95266--95290, 2024.

\bibitem[Wei et~al.(2023)Wei, Wang, Liu, Ding, and Zhang]{wei2023magicoder}
Yuxiang Wei, Zhe Wang, Jiawei Liu, Yifeng Ding, and Lingming Zhang.
\newblock Magicoder: Empowering code generation with oss-instruct.
\newblock \emph{arXiv preprint arXiv:2312.02120}, 2023.

\bibitem[Xu et~al.(2024)Xu, Lu, Shen, Wang, Wang, Mao, Xiong, and Yu]{agent-trek}
Yiheng Xu, Dunjie Lu, Zhenan Shen, Junli Wang, Zekun Wang, Yuchen Mao, Caiming Xiong, and Tao Yu.
\newblock Agenttrek: Agent trajectory synthesis via guiding replay with web tutorials.
\newblock \emph{arXiv preprint arXiv:2412.09605}, 2024.

\bibitem[Yen et~al.(2025)Yen, Gao, Hou, Ding, Fleischer, Izsak, Wasserblat, and Chen]{yen2025helmet}
Howard Yen, Tianyu Gao, Minmin Hou, Ke~Ding, Daniel Fleischer, Peter Izsak, Moshe Wasserblat, and Danqi Chen.
\newblock Helmet: How to evaluate long-context models effectively and thoroughly.
\newblock In \emph{The Thirteenth International Conference on Learning Representations}, 2025.

\bibitem[Zelikman et~al.(2024)Zelikman, Harik, Shao, Jayasiri, Haber, and Goodman]{quietstar}
Eric Zelikman, Georges Harik, Yijia Shao, Varuna Jayasiri, Nick Haber, and Noah~D. Goodman.
\newblock Quiet-star: Language models can teach themselves to think before speaking, 2024.
\newblock URL \url{https://arxiv.org/abs/2403.09629}.

\bibitem[Zhong et~al.(2024)Zhong, Cui, Guo, Liang, Lu, Wang, Saied, Chen, and Duan]{zhong2024agieval}
Wanjun Zhong, Ruixiang Cui, Yiduo Guo, Yaobo Liang, Shuai Lu, Yanlin Wang, Amin Saied, Weizhu Chen, and Nan Duan.
\newblock Agieval: A human-centric benchmark for evaluating foundation models.
\newblock In \emph{Findings of the Association for Computational Linguistics: NAACL 2024}, pages 2299--2314, 2024.

\end{thebibliography}

\clearpage
\beginappendix
\begin{table*}[ht]
\centering
\begin{tabular}{llcccc}
\toprule
\multirow{2.5}{*}{Family} & \multirow{2.5}{*}{\begin{tabular}[c]{@{}c@{}}Context Length \\ (tokens)\end{tabular}} & \multicolumn{4}{c}{Performance} \\
\cmidrule(lr){3-6}
& & Prolong & Raw-Repo & Repo2Agent & Repo2Agent-Search \\
\midrule
\multirow{3}{*}{NIAH-Multi} & 16384 & 99.40 & 99.50 & 99.60 & \textbf{99.70} \\
& 32768 & 98.40 & 99.00 & \textbf{99.20} & \textbf{99.20} \\
& 65536 & 66.20 & 76.30 & 68.70 & \textbf{80.40} \\
\midrule
\multirow{3}{*}{NIAH-Single} & 16384 & 100.00 & 100.00 & 100.00 & 100.00 \\
& 32768 & 100.00 & 99.90 & 99.90 & 99.90 \\
& 65536 & \textbf{92.10} & 89.90 & 90.30 & 91.30 \\
\midrule
\multirow{3}{*}{RULER-CWE} & 16384 & 80.40 & 85.00 & 79.90 & \textbf{87.30} \\
& 32768 & 27.60 & 34.60 & 33.20 & \textbf{42.30} \\
& 65536 & 0.30 & \textbf{6.50} & 0.40 & 1.60 \\
\midrule
\multirow{3}{*}{RULER-FWE} & 16384 & 92.70 & 92.10 & \textbf{95.10} & 92.20 \\
& 32768 & 87.70 & 88.10 & \textbf{90.10} & 86.00 \\
& 65536 & 73.70 & 70.30 & \textbf{79.20} & 67.90 \\
\midrule
\multirow{3}{*}{RULER-QA} & 16384 & 8.90 & 27.90 & \textbf{32.90} & 28.20 \\
& 32768 & 30.30 & 33.90 & 39.00 & \textbf{39.70} \\
& 65536 & 20.30 & \textbf{25.60} & 21.50 & 20.80 \\
\midrule
\multirow{3}{*}{RULER-VT} & 16384 & \textbf{99.40} & 99.00 & 99.00 & 98.20 \\
& 32768 & 95.20 & \textbf{96.00} & 95.30 & 94.10 \\
& 65536 & \textbf{20.50} & 14.40 & 18.60 & 16.60 \\
\midrule
\multirow{3}{*}{\textbf{Average}} & 16384 & 83.61 & 86.90 & \textbf{87.50} & 87.10 \\
& 32768 & 81.77 & 83.20 & 84.00 & \textbf{84.40} \\
& 65536 & 57.10 & 61.00 & 58.10 & \textbf{61.80} \\
\bottomrule
\end{tabular}
\caption{Results on Ruler. We average the results on NIAH-Multi-Key, NIAH-Multi-Value and NIAH-Multi-Query as NIAH-Multi-Multi. The results on RULER-QA-Hotpot and RULER-QA-Squad are averaged as RULER-QA.}
\label{tab:ruler}

\end{table*}
\FloatBarrier
\begin{table*}[ht]
\centering
\begin{tabular}{@{}l l S S S S@{}} 
\toprule
& & \multicolumn{4}{c}{Performance} \\ 
\cmidrule(l){3-6}
Category & Context Length & {Prolong} & {Raw-Repo} & {Repo2Agent} & {Repo2Agent-Search} \\ 
\midrule
ICL
& 16384 & 72.52 & 68.08 & 68.88 & \textbf{73.52} \\
& 32768 & 75.84 & 71.84 & 72.72 & \textbf{76.32} \\
& 65536 & \textbf{80.68} & 75.92 & 77.36 & 78.72 \\
\midrule
LongQA
& 16384 & 28.13 & 33.44 & \textbf{36.59} & 35.04 \\
& 32768 & 40.40 & 38.28 & \textbf{41.55} & 40.20 \\
& 65536 & \textbf{46.78} & 45.03 & 44.84 & 45.48 \\
\midrule
RAG
& 16384 & 64.17 & 63.88 & \textbf{64.46} & 64.08 \\
& 32768 & 63.33 & \textbf{63.67} & 63.13 & 63.25 \\
& 65536 & 56.00 & 57.42 & \textbf{57.67} & 55.58 \\
\midrule
Recall
& 16384 & 99.94 & 99.94 & 99.69 & 99.94 \\
& 32768 & 99.38 & 98.94 & 99.19 & \textbf{99.81} \\
& 65536 & 95.75 & 93.31 & 91.94 & \textbf{96.00} \\
\midrule
Rerank
& 16384 & 36.11 & 36.71 & \textbf{38.17} & 37.35 \\
& 32768 & 28.90 & 32.19 & 33.56 & \textbf{33.69} \\
& 65536 & 11.31 & 13.95 & \textbf{14.77} & 13.42 \\
\midrule
Avg
& 16384 & 60.17 & 60.41 & 61.56 & \textbf{61.99} \\
& 32768 & 61.57 & 60.98 & 62.03 & \textbf{62.65} \\
& 65536 & \textbf{58.10} & 57.13 & 57.32 & 57.84 \\
\bottomrule
\end{tabular}
\caption{Results on Helmet} 
\label{tab:helmet}
\end{table*}

\section{Detailed Results on Long-Context Benchmarks}
\label{app:detailed-long-bench}
\paragraph{Ruler}
Table~\ref{tab:ruler} presents the performance on the Ruler benchmark. The results highlight the robustness of agent-based training data at extreme context lengths.

\begin{itemize}
\item Superiority over Raw Code: Our proposed methods consistently outperform the internal \textit{Raw-Repo} baseline. For instance, at the 16k context length, \textit{Repo2Agent} achieves 87.50 compared to 86.90 for \textit{Raw-Repo}. This gap widens in specific tasks; in \textit{RULER-CWE} (32k), \textit{Repo2Agent-Search} scores 42.30, significantly outpacing \textit{Raw-Repo}'s 34.60.

\item Robustness at 64k: Performance stability at the maximum window size (64k) is a key differentiator. While the \textit{Prolong} baseline degrades to 57.10 and \textit{Raw-Repo} to 61.00, \textit{Repo2Agent-Search} maintains the highest robustness with an average score of \textbf{61.80}.

\item Complex Retrieval (NIAH): The benefits of agentic data are most pronounced in the \textit{NIAH-Multi} tasks, which require retrieving multiple pieces of scattered information---a process analogous to an agent locating dependencies across a file system. At 64k tokens, \textit{Repo2Agent-Search} achieves \textbf{80.40}, drastically outperforming \textit{Raw-Repo} (76.30) and establishing a massive lead over the \textit{Prolong} baseline (66.20).

\end{itemize}

\paragraph{Helmet}
The Helmet benchmark results (Table~\ref{tab:helmet}) further validate the efficacy of learning from trajectories, particularly in In-Context Learning (ICL) and Recall tasks.

\begin{itemize}
\item Consistent Gains over Raw-Repo: Across all context lengths (16k, 32k, and 64k), \textit{Repo2Agent-Search} consistently achieves a higher average score than the \textit{Raw-Repo} baseline. Notably, at 32k, our search-optimized model reaches \textbf{62.65} compared to 60.98 for raw code, demonstrating that the reasoning steps injected during training translate to better general understanding.

\item Recall and ICL Capabilities: \textit{Repo2Agent-Search} excels in tasks that mirror the ``Recall-Plan-Act'' loop of our synthetic agents. In the \textit{Recall} category, it outperforms both \textit{Raw-Repo} and the external \textit{Prolong} baseline at 32k (99.81) and 64k (96.00). Similarly, in \textit{ICL} (In-Context Learning), our method dominates at 16k and 32k, suggesting that observing the step-by-step history of code construction improves the model's ability to learn patterns from context.

\item Competitive Performance at Scale: While the \textit{Prolong} baseline shows strength in the \textit{LongQA} sub-task at 64k, leading to a slightly higher overall average (58.10), our method remains highly competitive (57.84) and crucially, still outperforms the \textit{Raw-Repo} baseline (57.13). To interpret these results correctly, it is necessary to distinguish the role of the baselines. While \textit{Prolong} serves as an external state-of-the-art reference, \textit{Raw-Repo} is the primary controlled ablation.
\end{itemize}

In summary, both benchmarks confirm that transforming raw repositories into dynamic agent trajectories is a more effective strategy than standard pre-training on static code, yielding models with sharper retrieval and reasoning capabilities.

\section{Case Study: Example of synthetic pretraining data}

\section{Case Study: Evolution of CoT}
\newtcolorbox{mycasebox}[2][]{%
    width=\textwidth,
    title={#2},                 
    fonttitle=\bfseries,        
    coltitle=white,             
    colbacktitle=black!75,      
    fontupper=\ttfamily\small, 
    colframe=black!75,
    boxrule=1pt,
    colback=black!5,
    nofloat,
    #1                          
}

\begin{mycasebox}[float*]{Original CoT in a Synthetic Agent Trajectory}
\label{box:case_original}
\begin{verbatim}
Now I understand that authentication state is managed via st.session_state 
by the authenticator. The 'status_2FA' flag is likely set in another 
page after 2FA setup, so I will proceed to implement the access control 
based on session state as described. I will now generate the full code 
for '2_Connect_Postgres_DB.py', including:
1. Page title
2. Loading authentication config from 'config.yaml'
3. Checking login and 2FA status
4. Initializing database connection with @st.cache_resource
5. Query execution with @st.cache_data
6. Displaying results in the required format
7. Error handling for database operations

I will write the code directly since no other function definitions are needed.
\end{verbatim}
\end{mycasebox}

\begin{mycasebox}[float*]{The CoT after 1st-round Search Optimization}
\label{box:case_round1}
\begin{verbatim}
Now I understand that the authentication state is managed through `st.session_state` 
by the authenticator. The `status_2FA` flag is likely set in a different 
page after 2FA setup, so I need to ensure the access control checks both 
`authentication_status` and `status_2FA` in the session state. I will 
proceed to implement the access control logic as described. Next, I will 
generate the full code for `2_Connect_Postgres_DB.py`, making sure to:
1. Set the page title.
2. Load the authentication configuration from `config.yaml`.
3. Check for valid login and 2FA status in the session state.
4. Initialize the PostgreSQL connection using `@st.cache_resource` for 
   one-time initialization.
5. Execute a query using `@st.cache_data` with a 5-minute cache TTL.
6. Display the query results in the specified format.
7. Include proper error handling for database connection and query execution.

I will now write the full code, ensuring all requirements are met.
\end{verbatim}
\end{mycasebox}

\begin{mycasebox}[float*]{The CoT after 2nd-round search optimization}
\label{box:case_round2}
\begin{verbatim}
Now I understand that the authentication state is managed through `st.session_state` 
by the authenticator. The `status_2FA` flag is likely set in a different 
page after 2FA setup, so I need to ensure the access control checks both 
`authentication_status` and `status_2FA` in the session state. I will 
proceed to implement the access control logic as described. Next, I will 
generate the full code for `2_Connect_Postgres_DB.py`, making sure to:

1. Set the page title to "Connect Postgres DB" to align with the user's 
   requirement.
2. Load the authentication configuration from `config.yaml` as it's used 
   to initialize the authenticator, which is crucial for the authentication flow.
3. Check for valid login and 2FA status in the session state. If either is 
   missing or not verified, display a warning and prevent further execution.
4. Initialize the PostgreSQL connection using `@st.cache_resource` for 
   one-time initialization, ensuring that the connection is not re-established 
   on every page rerun.
5. Execute a query using `@st.cache_data` with a 5-minute cache TTL, which 
   is specified in the requirement to cache query results.
6. Display the query results in the specified format, looping through the 
   fetched rows and formatting the output as `{row[0]} has a :{row[1]}:`.
7. Include proper error handling for database connection and query execution 
   to ensure the app handles failures gracefully.

I will now write the full code, ensuring all requirements are met, including 
environment variable loading, caching, and user access control. I will also 
make sure the code is modular, readable, and follows best practices for 
Streamlit app development.
\end{verbatim}
\end{mycasebox}

To provide a concrete illustration of how our LongCoT optimization refines the agent's reasoning, we present a case study tracking the evolution of a single Chain-of-Thought (CoT) step through multiple rounds of search. The task is to generate the code for a Python script, \code{2\_Connect\_Postgres\_DB.py}, which involves database interaction within a Streamlit application.

\paragraph{The Original CoT (Box \ref{box:case_original})} generated by the initial simulation is functional but generic. It outlines a correct but high-level plan, listing seven necessary steps, such as ``Checking login and 2FA status'' and ``Initializing database connection''. The reasoning is sparse, providing little detail on the ``how'' or ``why'' behind each step.

\paragraph{After the 1st Round of Optimization (Box \ref{box:case_round1})}, the CoT becomes more specific and technically precise. Key improvements include:
\begin{itemize}
    \item Explicit Variable Checks: It explicitly mentions the need to check both \code{authentication\_status} and \code{status\_2FA} in the session state, a detail absent in the original plan.
    \item Technical Justification: It begins to add rationale, explaining that \code{@st.cache\_resource} is for ``one-time initialization'' and specifying a ``5-minute cache TTL'' for \code{@st.cache\_data}. This demonstrates a deeper understanding of the tools being used.
\end{itemize}

\paragraph{The Final CoT after the 2nd Round (Box \ref{box:case_round2})} represents a significant leap in reasoning quality, transforming a simple checklist into a comprehensive implementation blueprint. The enhancements are substantial:
\begin{itemize}
    \item Detailed Rationale and User Intent: Each step is now accompanied by a rich explanation that links the action to a requirement. For example, it specifies the exact page title to ``align with the user's requirement'' and explains \textit{why} loading \code{config.yaml} is ``crucial for the authentication flow''.
    \item Elaboration on Edge Cases and Best Practices: The plan now includes explicit error handling logic (``display a warning and prevent further execution'') and implementation details (``looping through the fetched rows and formatting the output'').
    \item Holistic Project Awareness: The concluding thought expands beyond the immediate file, mentioning broader concerns like ``environment variable loading'', ``modularity, readability, and follows best practices for Streamlit app development''. This indicates a shift from a narrow, file-centric view to a more holistic, project-aware mindset.
\end{itemize}

This qualitative analysis empirically demonstrates that our search-based optimization does not simply rephrase CoTs. It systematically enriches the reasoning process, making it more detailed, explicit, and context-aware. This enriched reasoning, which more closely mirrors that of an expert developer, provides a much stronger learning signal for the model, which we believe is a key factor behind the performance improvements observed in our experiments.

\clearpage
\onecolumn
\section{Prompt}
\newtcblisting{markdownbox}[2][]{
    sharp corners,
    colframe=black!75,
    title={#2},                 
    fonttitle=\bfseries,        
    coltitle=white,             
    colbacktitle=black!75,      
    fontupper=\ttfamily, 
    colframe=black!75,
    boxrule=1pt,
    colback=black!5,
    listing only,
    listing options={
        breaklines=true,
        breakatwhitespace=true,
        basicstyle=\small\ttfamily,
        columns=fullflexible,
        keepspaces=true,
        showstringspaces=false,
        extendedchars=true,
    },
    breakable,
    enhanced,
    bottom=2pt,
    top=2pt,
    left=2pt,
    right=2pt,
    width=\linewidth 
}

\begin{markdownbox}{Prompt for Generating Main-agent Trajectory}
A github repo: $repo_code

The tree structure of repo: $file_tree.

Given the repo code and the tree structure of the repo, I want to use it to construct multi-agent synthetic data. The main agent needs to generate the implementation plan for the repo based on the detailed requirement document of the repo provided by the user, including the tree structure of the repo and the implementation order of files. It will also call sub-agents to realize the code generation of each file.
```
[
    {
        "role": "system_prompt",
        "content": "you are a helpful assistant. ... Show the sub-agent tool usage here.",
    },
    {
        "role": "user",
        "content": "A detailed requirement document for repo, but DO NOT mention implementation details of repo",
    },
    {
        "role": "gpt",
        "content": "tree structure of repo, implementation order of repo, call sub-agent to generate code for the first file",
        "tool-call": {
            "function_name": "code_generator",
            "arguments": {
                "requirement_for_repo": "requirement for repo",
                "tree_structure": "tree structure of repo",
                "file_name": "first_file.py",
                "file_path": "first_file.py",
                "requirement": "requirement for first_file.py",
            }
        }
    },
    {
        "role": "tool-response",
        "content": "return of function call",
    },
    {
        "role": "gpt",
        "content": "call sub-agent to generate code for the second file",
        "tool-call": {
            ...
        }
    },
    ... 
]
The Tool usage which should be put at the system prompt:
    Arguments of sub-agent:
    {requirement for repo, tree structure, file_name, file_path, requirement for file,}
    Return of sub-agent:
    {file_path has been generated successfully}

The memory of main agent should cover the planning of all the files in the repo, and call code-generator to generate all these files.
\end{markdownbox}

\begin{markdownbox}{Prompt for generating sub-agent trajectory}
        I have a GitHub repo, and I want to use it to construct multi-agent synthetic data. The main agent needs to generate the implementation plan for the repo based on the detailed requirement document of the repo provided by the user, including the repo's tree structure and the implementation order of files. It will also call sub-agents to realize the code generation of each file. The sub-agent requires information provided by the main agent, including the repo's requirement document, the repo's tree structure, the name and path of the code file that the sub-agent needs to generate, and the requirement description for this code file.

Your task is to generate a JSON list representing the simulated sub-agent's memory. This memory should chronicle the step-by-step thought process of creating a specific file from scratch, based on a user's requirement.

**Crucially, you are simulating the *creation* process, not explaining or refactoring existing code.** The agent you are simulating does not have access to the final source code at the beginning; it must figure out how to write it.

The format of the memory is as follows:
[
    {
        "role": "system_prompt",
        "content": "You are 'code_generator', an expert software engineer. \nYour goal is to implement robust, production-ready code from a given requirement.\n\nWorkflows:\n1. ANALYZE the file requirement and its place in the repo structure.\n2. IDENTIFY dependencies. If you need to use external classes/functions, use the `read` tool to check their definitions first.\n3. PLAN the implementation details (class structure, methods, logic).\n4. WRITE the code using the `write` tool.\n\nTools:\n- read(file_to_read): Returns the definition/signature of a file. Usage: When you need to understand how to invoke another module.\n- write(file_path, content): Writes the code to the file system.\n- final_answer(answer): Reports completion.",
    },
    {
        "role": "user",
        "content": "requirement for repo, tree structure, file_name, file_path, requirement for file",
    },
    {
        "role": "gpt",
        "content": "Here, the agent analyzes the requirement. It decides if external dependencies need to be checked based on the specific logic needed. It expresses curiosity or caution about specific interfaces it might need to interact with.",
        "tool-call": {
            "function_name": "read",
            "arguments": {
                "file_to_read": "file name",
            }
        }
    },
    {
        "role": "tool-response",
        "content": "the content of the file that was read",
    },
    {
        "role": "gpt",
        "content": "Here, the agent synthesizes the information from the requirement and any dependencies it read. It DOES NOT just list 'Plan: 1, 2, 3.' Instead, it narrates its engineering decisions, mentions specific variable names it *plans* to use, considers edge cases for the `file_name`'s logic, and explicitly reasons about how its planned implementation will satisfy the requirements.",
        "tool-call": {
            "function_name": "write",
            "arguments": {
                "file_path": "path of file",
                "content": "The source code that the agent decides to write."
            }
        }
    }
    ... (rest of the JSON structure)
]

To help you generate this simulated memory, you are provided with the following information. Use it as a guide to construct a realistic and accurate thought process.

*   **Information to construct the user prompt:**
    `$arguments_from_main_agent`
*   **The Golden Source Code for `$file_name` (The Goal):** This is the target code the simulated agent should ultimately produce. **You must not assume the agent has seen this code beforehand.** Use it as the "ground truth" to form a plausible thinking path that leads to this exact implementation.
    `$source_code`
*   **Source code of related files (Dependencies):** This is the content the agent will see when it uses the `read` tool on other files.
    `$related_source_code`

### CRITICAL INSTRUCTION: THOUGHT PROCESS DIVERSITY
The `content` fields in the "gpt" turns must contain **highly intelligent, specific, and varied** thought processes.
**STRICTLY AVOID** using the same template (e.g., "Okay, I have checked... Plan: 1. 2. 3.") for every file.

**Follow these guidelines to generate the thought process:**

1.  **Context-Driven Reasoning**:
    - If the **target** `$source_code` contains complex algorithms, the simulated thought process should focus on algorithmic efficiency and data structures.
    - If the **target** `$source_code` is a simple DTO or config file, the thought process should be brief and focused on correctness.
    - **Mention specific names**: The thought process MUST mention the actual class names, variable names, or function names found in the **target** `$source_code` and `$related_source_code` as part of its reasoning and planning.

2.  **Dependency Logic**:
    - When simulating a `read` call: Explain *specifically* what the agent is looking for (e.g., "I need to see if the `User` class has a `get_id` method or just a public `id` field before I can implement the logic that uses it.").
    - After simulating a `read`: The agent should react to the content found (e.g., "Ah, I see `User`'s constructor requires a positional argument, not a keyword argument. I'll make sure to call it correctly in my implementation.").

# Output Format
Return strictly a JSON list representing the memory.
\end{markdownbox}

\begin{markdownbox}{{Prompt for Optimizing CoT}}
You are an expert software engineer. Your task is to simulate the human reasoning process required to solve a programming problem.

**The Goal:**
You need to rewrite a specific part of a reasoning chain (the "Target Block"). The goal is to make the reasoning logic more precise, detailed, and aligned with the correct solution, WITHOUT breaking the narrative flow.

**Input Data:**
1. **Reference Source Code:** (The correct answer, for your understanding ONLY)
{}

2. **Full Reasoning Context:** (The story so far)
{}

3. **Target Block to Rewrite:** (The weak step needs replacing)
<replace>
{}
</replace>

**CRITICAL INSTRUCTIONS (Read Carefully):**

1.  **The "Time Travel" Rule:**
    You must act as if you are solving this problem *for the first time*. You do NOT know the final code yet; you are currently deriving it.
    * **STRICTLY FORBIDDEN:** Do not mention "Reference Code", "Provided Solution", or "Ground Truth".
    * **CORRECT APPROACH:** Instead of saying "The reference code uses a HashMap...", say "I think a HashMap would be the best data structure here because..."

2.  **The "Invisible Stitch" Rule:**
    Your output will be copy-pasted directly into the original text to replace the old block. It must fit perfectly.
    * **STRICTLY FORBIDDEN:** Do not verify or announce the correction. Never use phrases like "In this refinement...", "Correcting the previous step...", "Here is the better reasoning...", or "Let's refine this".
    * **CORRECT APPROACH:** Just write the thought process directly. Start immediately with "I need to analyze...", "Next, I will...", etc.

3.  **Tone & Style:**
    * Use **First-Person Singular** ("I check...", "I decide...").
    * Use **Present Tense** (Reasoning happens *now*).
    * Be technical, precise, and deductive.

**Your Workflow:**
1.  **Analyze (`<think>` tags):**
    * Briefly analyze the Reference Code to understand the *correct* logic.
    * Identify why the original `Target Block` was weak or incorrect.
    * Plan the logic steps needed to bridge the gap.
2.  **Generate (`<refine>` tags):**
    * Write the purely deductive thought process.
    * Ensure it starts and ends in a way that connects with the surrounding text in `reasoning_chain`.

Now, generate the replacement block.

<think>
[Your analysis of the gap between the reasoning and the code]
</think>

<refine>
[The seamless, first-person reasoning stream ONLY]
</refine>
\end{markdownbox}

\end{document}